\documentclass[aps,prb,twocolumn,showpacs]{revtex4-1}
\usepackage{graphicx}
\usepackage{dcolumn}
\usepackage{bm}
\usepackage{amsmath}
\usepackage{color}
\usepackage[normalem]{ulem}

\begin{document}

\newcommand{\bra}[1]{\left\langle#1\right|}
\newcommand{\ket}[1]{\left|#1\right\rangle}
\newcommand{\bracket}[2]{\big\langle#1 \bigm| #2\big\rangle}
\newcommand{\Tr}{{\rm Tr}}
\renewcommand{\Im}{{\rm Im}}
\renewcommand{\Re}{{\rm Re}}
\newcommand{\ef}{{\epsilon_{\rm F}}}
\newcommand{\MC}[1]{\mathcal{#1}}
\newcommand{\pp}{{\prime\prime}}
\newcommand{\ppp}{{\prime\prime\prime}}
\newcommand{\pppp}{{\prime\prime\prime\prime}}

\title{Kondo physics of the Anderson impurity model by Distributional Exact Diagonalization}

\author{S. Motahari} 
\affiliation{Max-Planck-Institut f\"ur Mikrostrukturphysik, Weinberg 2, 06120 Halle, Germany}
\author{R. Requist} 
\affiliation{Max-Planck-Institut f\"ur Mikrostrukturphysik, Weinberg 2, 06120 Halle, Germany}
\author{D. Jacob}\email{djacob@mpi-halle.mpg.de}
\affiliation{Max-Planck-Institut f\"ur Mikrostrukturphysik, Weinberg 2, 06120 Halle, Germany}

\date{\today} 

\begin{abstract}
  The Distributional Exact Diagonalization (DED) scheme is applied to the description of Kondo 
  physics in the Anderson impurity model. DED maps Anderson's problem of an interacting impurity 
  level coupled to an infinite bath onto an ensemble of finite Anderson models, each 
  of which can be solved by exact diagonalization. An approximation to the self-energy of the 
  original infinite model is then obtained from the ensemble averaged self-energy. 
  Using Friedel's sum rule, we show that the particle number constraint, a central ingredient of 
  the DED scheme, ultimately imposes Fermi liquid behavior on the ensemble averaged self-energy,
  and thus is essential for the description of Kondo physics within DED.
  Using the Numerical Renormalization Group (NRG) method as a benchmark, we show that DED yields excellent
  spectra, both inside and outside the Kondo regime for a moderate number of bath sites. 
  Only for very strong correlations ($U/\Gamma\gg10$) does the number of bath sites needed to achieve
  good quantitative agreement become too large to be computationally feasible.
\end{abstract}

\maketitle

\section{Introduction}
\label{sec:intro}

The Anderson impurity model (AIM)\cite{Anderson:PR:1961} plays a central role in the understanding of 
one of the most intriguing many-body phenomena, the Kondo effect,\cite{Hewson:book:1997} 
and is also at the heart of Dynamical Mean-Field Theory (DMFT).\cite{Metzner:PRL:1989,Georges:PRB:1992,Georges:RMP:1996,Kotliar:RMP:2006}
The Numerical Renormalization Group method\cite{Bulla:RMP:2008} solves the model exactly, 
but is computationally very demanding and unable to make use of the strongest form of parallelization. 
Another numerically exact method for solving the AIM is the Continuous-time Quantum Monte Carlo (CTQMC) algorithm,\cite{Gull:RMP:2011}
which can be parallelized efficiently, but has the disadvantage of working in imaginary time. 
The necessary analytical continuation back to the real axis brings about artifacts in the spectral function.
Another serious drawback of CTQMC is its restriction to relatively high temperatures, making this approach
of limited use for the study of low-temperature phenomena such as the Kondo effect.

A number of approximate methods for solving the Anderson model exist as well. 
The Non-Crossing Approximation (NCA)\cite{Grewe:PRB:1981,Coleman:PRB:1984} and 
One-Crossing Approximation (OCA),\cite{Pruschke:ZPB:1989,Haule:PRB:2001} for example,
consist in a diagrammatic expansion around the atomic limit, summing only a subset of
diagrams to infinite order. Both NCA and OCA yield qualitatively correct spectra for
not too low temperatures. While the simpler NCA strongly underestimates the width of 
the Kondo peak, the vertex corrections within OCA lead to a quantitatively correct
estimate of the Kondo scale. At lower temperatures, both NCA and OCA show spurious 
non-Fermi liquid behavior, leading to artifacts in the spectra.\cite{Costi:PRB:1996,Grewe:JPCM:2008}
Many other approximate schemes for solving the AIM 
exist,\cite{Yosida:PTP:1970,Read:JPhysC:1983,Logan:JPCM:1998,Hewson:JPCM:2001,Feng:JPCM:2011} 
though all are burdened with some kind of limitation.

Common to most approximation schemes is the solution of the infinite AIM, consisting of an 
impurity level coupled to an infinite and continuous bath representing a conduction electron band. 
A different route is to replace the infinite AIM by a finite one that can then be solved by numerical 
diagonalization.\cite{Caffarel:PRL:1994,Liebsch:JPCM:2012} 
The infinite and continuous conduction electron bath is approximated by a finite number of discrete bath levels. 
When this approach is adopted in DMFT as an impurity solver, it yields thermodynamic and static 
quantities in very good agreement with e.g.~numerically exact CTQMC but often leads 
to artifacts in the spectral functions stemming from finite size effects. Especially in the Kondo 
regime, the discrete nature of the conduction electron bath in the exact diagonalization approach 
seriously compromises the correctness of the impurity density of states, a key observable\cite{Ujsaghy:PRL:2000} 
in the scanning tunneling spectroscopy of surface Kondo systems such as Ce on silver\cite{Li:PRL:1998} 
or Co on gold\cite{Madhavan:S:1998} and 
copper\cite{Manoharan:N:2000,Knorr:PRL:2002,Wahl:PRL:2004,Neel:PRL:2007,Vitali:PRL:2008,Surer:PRB:2012,Jacob:JPCM:2015,Baruselli:PRB:2015,Frank:PRB:2015} surfaces.

Recently, Granath and Strand have proposed a novel method for solving the AIM that overcomes 
the problem of discretization artifacts. 
The Distributional Exact Diagonalization (DED) approach\cite{Granath:PRB:2012,Granath:PRB:2014}
maps the infinite Anderson model onto an \textit{ensemble} of finite Anderson models instead 
of a single effective finite Anderson model.
The ensemble average of the self-energies of the finite Anderson models provides a smooth
approximation to the self-energy of the original infinite Anderson model that is also free of finite-size artifacts.
An advantage of the DED method in comparison with NRG is its straightforward and efficient 
large-scale parallelization. Different strategies for improving direct diagonalization methods have been proposed recently. 
In one, a careful selection of basis states makes it possible to include a large number of bath levels.\cite{Lu:PRB:2014}  
In another, the parameters of an effective finite Anderson model are variationally optimized.\cite{Schuler:PRB:2015}

Here we show that the DED approach gives an excellent description of the Anderson model inside and outside the Kondo regime, 
except for very strong correlations. We find that already for a very small number of 1-2 bath sites, the spectra 
are in good qualitative agreement with exact spectra calculated by NRG. 
For a moderate number of 5-7 bath sites the agreement becomes excellent, also with regard to the width of the Kondo peak.
Only for very strong correlation strength ($U/\Gamma\gg10$) does the number of bath sites necessary to obtain quantitative results become
computationally prohibitive due to the exponential growth of the Kondo screening cloud with correlation strength.

The paper is organized as follows. In Sec.~II we first review the DED method, originally introduced 
by Granath and Strand, and then elucidate the role of the particle number constraint that is needed to make the method work. 
In Sec.~III we apply the DED method to the single-orbital AIM, both in the particle-hole (ph) 
symmetric case (Sec.~III.~A) and in the presence of asymmetry (Sec.~III.~B). 
Finally, in Sec.~IV we conclude the paper with a discussion of the results and a perspective on using DED 
for more general types of Anderson impurity models.

\section{Method}
\label{sec_method}

\subsection{Review of the DED Algorithm}
\label{sub_review_ded}

We consider the AIM of a single interacting impurity level coupled to an infinite bath of conduction electrons:
\begin{eqnarray}
  \label{eq:AIM}
  H &=& \epsilon_d n_d + U n_{d\uparrow}n_{d\downarrow} 
  +\sum_{\sigma,k}\epsilon_k\,c^{\dagger}_{k\sigma}c_{k\sigma}
  \nonumber\\
  &&+\sum_{\sigma,k} V_k \, (d^{\dagger}_{\sigma}c_{k\sigma} + c^{\dagger}_{k\sigma}d_{\sigma}) 
\end{eqnarray}
with $d_\sigma$ ($d_\sigma^\dagger$) the annihilation (creation) operator for the impurity level $d$ and spin $\sigma$,
$c_{k\sigma}$ ($c_{k\sigma}^\dagger$), the  annihilation (creation) operators for bath levels $k$ and spin $\sigma$,
$n_{d\sigma}=d_\sigma^{\dagger}d_\sigma$, $n_d=\sum_\sigma n_{d\sigma}$, $\epsilon_d$ the bare impurity level energy, 
$U$ the on-site Coulomb repulsion at the impurity, $\epsilon_k$ the band energy of conduction electrons,
and $V_k$ the coupling between the impurity level $d$ and conduction electron $k$.
The chemical potential $\mu$ is assumed to be zero throughout the paper.

\begin{figure*}
  \includegraphics[width=1.0\textwidth]{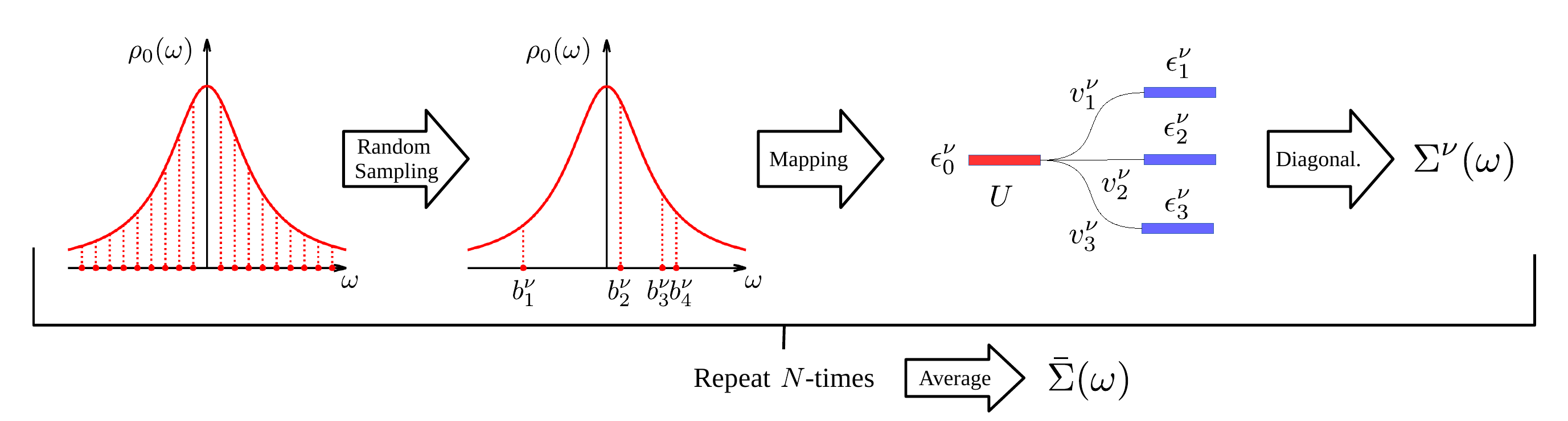}
  \caption{\label{fig:method_scheme}
    Schematic representation of the DED method. The non-interacting density of the impurity level $\rho_0(\omega)$
    is interpreted as a probability distribution for the poles of the non-interacting Green's function $G_0(\omega)$.
    A finite number of poles $b_i^\nu$ is then generated randomly according to the distribution $\rho_0(\omega)$.
    The selected $n$ poles uniquely define a finite Anderson model $H^\nu$ with $n-1$ bath sites. Diagonalization of
    $H^\nu$ yields the self-energy $\Sigma^\nu(\omega)$ corresponding to the finite Anderson model. This process is
    repeated many times ($N$). An approximation to the self-energy of the original infinite Anderson model 
    (\ref{eq:AIM}) is obtained from the ensemble average of the self-energies of the finite Anderson model 
    samples (\ref{eq:average}).
  }
\end{figure*}

The general idea of the DED approach is to map the infinite Anderson model to an ensemble of relatively small 
\emph{finite} Anderson models that can be diagonalized exactly. Our starting point is the non-interacting 
retarded Green's function:
\begin{equation}
  \label{eq:G0}
  G_0(\omega) = \frac{1}{\omega^+-\epsilon_d-\Sigma_0-\Delta(\omega)}
\end{equation}
where $\Delta(\omega)$ is the hybridization function $\Delta(\omega)=\sum_k\frac{(V_k)^2}{\omega^+-\epsilon_k}$,
describing the renormalization (real part) and broadening (imaginary part) of the impurity level
due to the coupling to the conduction electron bath. For all calculations presented here
we assume a flat hybridization function, i.e.~$\Delta(\omega)=-i\Gamma$ (wide band limit), but the 
approach is not limited in that respect. 

The parameter $\Sigma_0$ can be understood as an effective one-body potential 
for the non-interacting reference system. Its exact role will be elucidated 
later in the context of the constraint (see Sec.~\ref{sub_constraint}).
Anticipating our later discussion, we mention here that Fermi liquid theory considerations
suggest that $\Sigma_0$ should be the real part of the interacting self-energy at the Fermi level.
In including $\Sigma_0$ already at this stage, and interpreting it as an effective one-body
potential, our approach deviates somewhat from the one originally proposed by Granath and Strand~\cite{Granath:PRB:2012}
(see Sec.~\ref{sub_constraint} for a detailed discussion).

Next, $G_0$ is represented by a large number $M$ of poles $b_i$ on the real axis, thereby effectively 
discretizing the conduction electron bath:
\begin {equation}
  \label{eq:GF0}
  G_0(\omega)=\sum_{i=1}^{M}\frac{{a}_i}{\omega^+-b_i} {.} 
\end {equation}
Here $a_i$ are the residues corresponding to the poles $b_i$ which have to be normalized according 
to $\sum{{a}_i}=1$. We then divide the poles into $N$ groups of size $n$ ($Nn=M$):
\begin {equation}
  G_0(\omega)=\frac{1}{N}\sum_{\nu=1}^{N}\sum_{i=1}^{n}\frac{a_i^\nu}{\omega^+-b_i^\nu}=\frac{1}{N}\sum_{\nu}G_0^{\nu}(\omega) {,}
\end {equation}
where $n$ is a relatively small integer number that ultimately determines the size of the finite AIM, 
and $N$ the number of finite-size Anderson model samples in the ensemble.
Now the residues in \emph{each group} have to be normalized according to 
$\sum_{i=1}^n{{a}_i^\nu}=1$ for all $\nu=1\ldots{N}$. 

The poles representing $G_0(\omega)$ 
are generated randomly using the non-interacting spectral density 
$\rho_0(\omega)=-\Im[G_0(\omega)]/\pi$ as the probability distribution. 
Each set $\nu$ of $n$ such randomly chosen poles then uniquely defines the non-interacting part
of a finite-size ($n$ sites) Anderson model:
\begin{eqnarray}
 H_0^{\nu} &=&\epsilon_0^{\nu}\sum_{\sigma}d^{\dagger}_{\sigma}d_{\sigma}
 +\sum_{\sigma,k=1}^{n-1}V_k^{\nu}(d^{\dagger}_{\sigma}c_{k\sigma} +c^{\dagger}_{k\sigma}d_{\sigma}) 
 \nonumber\\
 &&+\sum_{\sigma,k=1}^{n-1}\epsilon_k^{\nu}\,c^{\dagger}_{k\sigma}c_{k\sigma} {.}
\end{eqnarray}
The mapping from the set of poles to the parameters of the finite Anderson model is achieved 
by equating $G_0^{\nu}(\omega)$ and the impurity Green's function (GF) corresponding to $H_0^{\nu}$:
\begin{equation}
  \sum_{i=1}^{n}\frac{a_i^{\nu}}{\omega^+-b_i^{\nu}}=\left(\omega^+-\epsilon_0^{\nu}-\sum_{k=1}^{n-1}\frac{(V_k^{\nu})^2}{\omega^+-\epsilon_k^{\nu}}\right)^{-1} {,}
\end{equation}
where the residues are taken to be constant with $a_i^\nu=1/n$. 
Note that since the poles $b_i^\nu$ are chosen 
to be distributed randomly according to the probabilities $\rho_0(b_i^\nu)$, the seemingly reasonable 
choice $a_i^\nu\sim\rho_0(b_i^\nu)$ for the residues is actually wrong as it would lead to a sampled 
non-interacting DOS different from $\rho_0(\omega)$.
The bath energy levels $\epsilon_k^\nu$ can now be found from the roots of $G_0^{\nu}$,
\begin{equation}
  \label{eq:ek}
  G^\nu_0(\omega=\epsilon_k^\nu)=0 \hspace{1ex} \forall \; k=1,\ldots,n-1,
\end{equation}
while the hoppings $V_k^\nu$ between the impurity and the bath levels are
obtained from the derivative of $G_0^{\nu}$ at the bath level energies as
\begin{equation}
  \label{eq:Vk}
  \left.\frac{dG_0^\nu}{d\omega}\right|_{\epsilon_k^\nu}=-\frac{1}{(V_k^\nu)^2} \hspace{1ex} \forall \; k=1,\ldots,n-1 {.}
\end{equation}
Finally, the impurity level energy is obtained from the mean value of sampled poles:
\begin{equation}
  \label{eq:e0}
  \epsilon_0^\nu=\sum_{i=1}^n a_i^\nu b_i^\nu = \frac{1}{n} \sum_{i=1}^n b_i^\nu
\end{equation}
In the next step, the interacting finite Anderson model is obtained by adding the interaction
part, and, importantly, subtracting out the effective one-body potential $\Sigma_0$, to 
avoid double counting of interactions:
\begin{equation}
 H^{\nu} = H_0^{\nu} + U n_{d\uparrow}n_{d\downarrow} - \Sigma_0 n_d
\end{equation}
Hence we see that $\Sigma_0$ does not really play a role yet. 
The role of $\Sigma_0$ will become clear later
in the context of the constraint (see Sec.~\ref{sub_constraint}).
For later convenience we also define the bare impurity level $\epsilon_d^\nu=\epsilon_0^\nu-\Sigma_0$
of the finite model. Note that $\epsilon_d^\nu\rightarrow\epsilon_d$ in the limit of $n\rightarrow\infty$.

The finite Anderson model Hamiltonian $H^{\nu}$ is now diagonalized numerically. This yields
the many-body eigenstates $\ket{m^\nu}$ and corresponding eigenenergies $E_m^\nu$. 
The GF for the impurity level is then obtained from the Lehmann representation:
\begin{eqnarray}
  \label{eq:Lehmann}
  G_\sigma^\nu(\omega) &=& 
  \sum_{m}\frac{|\langle m^\nu|d_\sigma|0^\nu\rangle|^2}{\omega^++E_m^\nu-E_0^\nu} + \sum_{m}\frac{|\langle m^\nu|d^\dagger_\sigma|0^\nu\rangle|^2}{\omega^++E^\nu_0-E_m^\nu} 
  \nonumber\\
\end{eqnarray}
where $\ket{0^\nu}$ and $E_0^\nu$ denote the ground state and corresponding ground state energy.
\footnote{
  In the case of a degenerate ground state the GF would be obtained from the corresponding 
  ensemble average over the ground state manifold. Note, however, that the particle 
  constraint discussed in Sec.~\ref{sub_constraint} ensures that the ground state 
  is actually a singlet state. 
}
The corresponding self-energy of the finite Anderson model is
\begin{equation}
  \label{eq:sample_selfenergy}
  {\Sigma}^{\nu}_\sigma(\omega) = (G_0^{\nu}(\omega))^{-1} - (G^{\nu}_\sigma(\omega))^{-1} + \Sigma_0 
\end{equation}
This process of generating finite Anderson model Hamiltonians $H^\nu$ and calculating their self-energies ${\Sigma}^{\nu}_\sigma$
is repeated $N$ times. 
Finally, an approximation to the self-energy of the original \emph{infinite} Anderson model is obtained from the ensemble average
\begin{equation}
  \label{eq:average}
  \bar\Sigma_\sigma(\omega) = \frac{1}{N}\sum_{\nu=1}^N \Sigma^{\nu}_\sigma(\omega)
\end{equation}
An approximation to the corresponding interacting GF is obtained from $G_\sigma(\omega)=(\omega^+-\epsilon_d-\bar\Sigma_\sigma(\omega)-\Delta(\omega))^{-1}$.
As observed by Granath and Strand, obtaining an approximation to the GF of the infinite Anderson model 
by directly averaging the $G^{\nu}_\sigma(\omega)$ is not an option, since the sample-averaged interacting and 
non-interacting GFs $\bar{G}_\sigma(\omega)=\frac{1}{N}\sum_{\nu}G^\nu_\sigma(\omega)$ and 
$\bar{G}_0(\omega)=\frac{1}{N}\sum_{\nu}\bar{G}^\nu_\sigma(\omega)$,
respectively, do not form a proper pair of interacting and non-interacting 
GFs connected by the Dyson equation.\cite{Granath:PRB:2012}
Fig.~\ref{fig:method_scheme} shows a schematic representation summarizing the main steps of the DED procedure.

\subsection{Role of the constraint}
\label{sub_constraint}

\begin{figure}
  \includegraphics[width=\linewidth]{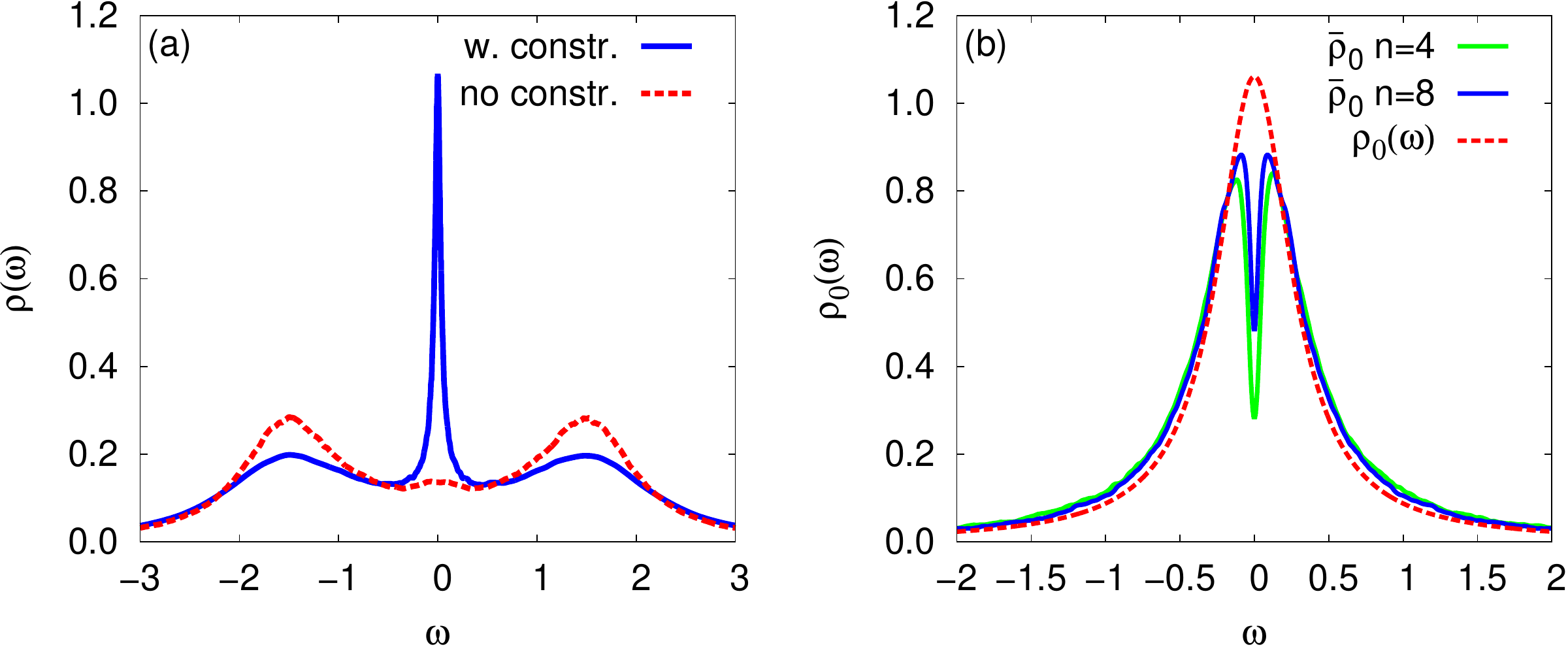} 
  \caption{\label{fig:constraint}
    Effect of constraint: (a) Comparison of spectra calculated with (blue line) and without (red dashed 
    line) imposing the particle number constraint for $n=4$ sites, $\Gamma=0.3$, $U=3$, $\epsilon_d=-U/2$.
    (b) Comparison of sampled non-interacting DOS $\bar\rho_0(\omega)=-\Im~\bar{G}_0(\omega)/\pi$ for different 
    number of sites with original Lorentzian non-interacting DOS $\rho_0(\omega)$.
  }
\end{figure}

Granath and Strand found that in order to obtain valid spectra not all randomly generated 
Anderson models can be accepted. 
As can be seen in Fig.~\ref{fig:constraint}a (red dashed line), the Kondo peak is practically
non-existent and the Hubbard side peaks are overestimated when all randomly generated 
finite Anderson model samples contribute equally.
In order to deal with this problem, Granath and Strand introduced a constraint
comparing the number of particles in the interacting and non-interacting systems.
More precisely, a sample $\nu$ is only accepted if 
\begin{equation}
  \label{eq:constraint}
  N^\nu \stackrel{(!)}{=} N_0^\nu
\end{equation}
where $N^\nu$ is the number of particles of the ground state of the interacting model $H^\nu$
and $N_0^\nu$ that of the non-interacting model $H_0^\nu$. As can be seen in Fig.~\ref{fig:constraint}a
(blue line), applying the constraint indeed recovers the full height of the Kondo peak
at the Fermi level and lowers the Hubbard side peaks. The effect of the constraint on
the sampled non-interacting DOS is to deplete the DOS around the Fermi level as can be seen in 
Fig.~\ref{fig:constraint}b. As the number of sites $n$ increases the effect of the constraint becomes 
smaller.

In order to elucidate the role played by the constraint, we now consider Friedel's sum rule for 
the Anderson model\cite{Langer:PR:1961,Langreth:PR:1966,Hewson:book:1997} which relates the 
impurity charge $n_{\rm imp,\sigma}$ to the scattering phase shift at the Fermi level. For a finite
Anderson model sample $\nu$ we can write:
\begin{equation}
\label{eq:friedel}
  n_{\rm imp,\sigma}^\nu = \eta_\sigma^\nu(\epsilon_{\rm F}) / \pi
\end{equation}
where the scattering phase shift $\eta^\nu_\sigma$ is given by
\begin{equation}
\label{eq:eta}
 \eta_\sigma^\nu(\omega) = \frac{\pi}{2} 
 -\tan^{-1}\left( \frac{\omega - \epsilon^\nu_d - \Re\Sigma^\nu_\sigma(\omega) - \Re\Delta^\nu(\omega) }{ \Im\Sigma^\nu_\sigma(\omega)+\Im\Delta^\nu(\omega)} \right)
\end{equation}
As we are considering a \emph{finite} Anderson model, the hybridization function
\begin{equation}
  \Delta^\nu(\omega) = \sum_k \frac{|V_k^\nu|^2}{\omega^+-\epsilon_k^\nu}
\end{equation}
which describes the coupling of the impurity level with the bath levels,
is the sum of a finite number of poles, and thus non-constant by construction.
In this case the impurity charge $n_{\rm imp,\sigma}^\nu$ comprises not only the 
impurity level occupancy $n_{d,\sigma}^\nu$ but also the additional scattering induced 
charge $\delta{n}_{\rm imp,\sigma}$:
\begin{eqnarray}
  \lefteqn{n_{\rm imp,\sigma}^\nu = n_{d,\sigma}^\nu + \delta{n}^\nu_{\rm imp,\sigma}}
  \nonumber\\
  && \hspace{2ex} = -\int_{-\infty}^{\ef} \frac{d\omega}{\pi}\, 
  \Im\left( G_\sigma^\nu(\omega) + \sum_k \left(g_k^\nu(\omega)\right)^2 T^{\nu}_{k\sigma}(\omega) \right)
  \hspace{3ex}
\end{eqnarray}
where $g_k^\nu(\omega)=1/(\omega^+-\epsilon_k^\nu)$ is the propagator for the \emph{isolated} (i.e. not connected to the impurity)
bath-level $k$, and $T^{\nu}_{k\sigma}(\omega)=V_k^\nu G_\sigma^\nu(\omega) V_k^\nu$ is the scattering T-matrix. 

The total number of electrons $N^\nu$ for sample $\nu$ is given by the sum of the impurity charge 
$n_{\rm imp,\sigma}^\nu$ and the occupation of the the \emph{isolated} bath levels $n_{\rm bath}^\nu=-\Im\int_{-\infty}^\ef{d\omega}\sum_k{g_k(\omega)}/\pi$.
Since the occupation of the \emph{isolated} bath levels is the same in the interacting and non-interacting system,
the particle constraint ultimately imposes that the impurity charge, and in turn the phase shifts, are the same 
for the interacting and non-interacting models:
\begin{equation}
  \label{eq:constraint2}
  n_{\rm imp,\sigma}^\nu = \frac{\eta^\nu_\sigma(\ef)}{\pi} \; \stackrel{\rm constr.}{=} \; n_{\rm imp,\sigma,0}^\nu = \frac{\eta_{\sigma,0}^\nu(\ef)}{\pi}
\end{equation}
where the phase shift of the non-interacting system is given by
\begin{equation}
  \label{eq:eta0}
  \eta_{0,\sigma}^\nu(\omega) = \frac{\pi}{2} - \tan^{-1}\left( \frac{\omega - \epsilon^\nu_0 - \Re\Delta^\nu(\omega) }{ \Im\Delta^\nu(\omega)} \right)
\end{equation}
It is here that the effective potential $\Sigma_0$ included in the non-interacting GF (\ref{eq:G0}) 
enters in the constraint: Since $\epsilon_0^\nu=\epsilon_d^\nu+\Sigma_0$ it determines the phase shift $\eta_{0,\sigma}^\nu(\ef)$
and consequently the impurity charge $n_{{\rm imp},\sigma,0}^\nu$ of the non-interacting reference system.

Hence the constraint guarantees that only self-energies $\Sigma^\nu_\sigma(\omega)$ which have the same phase shift 
as the corresponding non-interacting model contribute to the ensemble average (\ref{eq:average}). 
A closer look at the phase shifts $\eta^\nu_\sigma$ 
and $\eta^\nu_{0,\sigma}$ of individual finite Anderson model samples $\nu$ further reveals that the constraint 
really establishes a 1:1 correspondence between the excitations of the interacting Hamiltonian $H^\nu$ and the 
corresponding non-interacting one $H^\nu_0$, as required by Fermi liquid theory (see App.~\ref{app_constraint} for details).
When the constraint is not fulfilled, the 1:1 correspondence with the non-interacting 
system cannot be established, because the ground state of the interacting system has an odd number of electrons
($N^\nu=N^\nu_0\pm1$) and thus is a doublet state ($S=1/2$), while the non-interacting system must have an 
even number of electrons (single-particle levels are either doubly occupied or unoccupied), and thus has a 
singlet ground state ($S=0$). 
Thus the constraint ultimately enforces that individual Anderson model samples contributing to 
the self-energy average (\ref{eq:average}) comply with Nozieres' Fermi liquid picture \cite{Nozieres:JLTP:1974} 
of the Kondo effect in the strong coupling regime: the impurity spin locks into a total spin singlet state 
with a few conduction electron bath levels, and the remaining conduction electrons interact weakly 
with the singlet state, thus leading to Fermi liquid behavior.
Since Friedel's sum rule is directly related to the height of the Kondo peak at the Fermi energy,
the particle constraint ultimately leads to the recovery of the unitary limit for the interacting spectral function, 
and consequently to the recovery of Fermi-liquid behavior.

The interpretation of the constraint as a sample-wise enforcement of Fermi liquid behavior suggests that 
the parameter $\Sigma_0$ should be interpreted as an effective one-body potential that can be identified 
with the real part of the (yet to be determined) many-body self-energy: 
\begin{equation}
  \label{eq:sigma0}
  \Sigma_0 \equiv \Re\bar\Sigma(\ef)
\end{equation}
This conjecture can be further justified by considering the exact limit of the DED approach:
taking the number of poles $n\rightarrow\infty$, the original infinite Anderson model is
recovered. Since now there is only one sample, the constraint must be fulfilled for this
one sample, hence the phase shift of the interacting and corresponding non-interacting 
model must match exactly, leading to:
\begin{equation*}
  \tan^{-1}\left( \frac{\epsilon_d + \Re\Sigma(\ef) - \ef}{\Gamma} \right) 
  \stackrel{(!)}{=} \tan^{-1}\left( \frac{\epsilon_d + \Sigma_0 - \ef}{\Gamma} \right) 
\end{equation*}
which implies (\ref{eq:sigma0}).

Since the self-energy itself is not known prior to the calculation, $\Sigma_0$ has to be determined
self-consistently, starting with some initial guess for $\Sigma_0$, for example the Hartree shift 
$\Sigma_0\equiv{Un_d/2}$ with $n_d$ being the Hartree-Fock occupancy.
This is where our approach slightly differs from the one originally proposed by Granath and Strand,
which takes $\Sigma_0$ as an adjustable parameter to be fixed by demanding that the interacting and 
non-interacting impurity occupancy $n_d$ be the same.

\section{Results}
\label{sec_results}

In the following we present results for the AIM described by eq.~(\ref{eq:AIM}), assuming a constant hybridization function 
$\Delta(\omega)=-i\Gamma$. The non-interacting density of states $\rho_0(\omega)$ is thus a Lorentzian centered at $\epsilon_d+\Sigma_0$
of width $2\Gamma$. To resolve the interacting spectral functions we use a logarithmic mesh, and a frequency dependent Lorentzian
broadening scheme where an imaginary part proportional to the frequency is added to the frequency argument in the denominators
of the Green's functions, i.e. $\omega^+=\omega+i\eta_1\cdot|\omega|$ with $\eta_1=0.02$.  
The NRG calculations were performed with the NRG Ljubljana code \cite{Zitko}, using the $z$-averaging technique \cite{Zitko:PRB:2009} with $z=64$.
For all calculations, we set the conduction band half-width to $D=10$, the logarithmic discretization parameter to $\Lambda=2$ and determined the 
number of states kept at each iteration by an energy cutoff of $10\omega_N$ ($\omega_N$ is the characteristic energy scale of iteration $N$); the 
maximum number of states kept was 6,600 counting multiplicities.
Log-Gaussian broadening\cite{Bulla:PRB:2001} was used in the calculation of the spectral functions with a width parameter of $\alpha=0.2$ for 
the asymmetric AIM.  For the symmetric AIM, $\alpha$ was varied between $0.15$ for small $\Gamma$ and $0.35$ for large $\Gamma$. 

\subsection{Symmetric Anderson model}
\label{sub_symm_aim}

\begin{figure}[h]
  \begin{tabular}{cc}
  \includegraphics[width=0.48\linewidth]{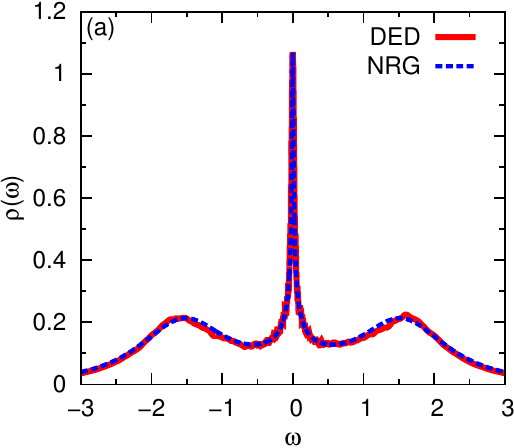} &
    \includegraphics[width=0.48\linewidth]{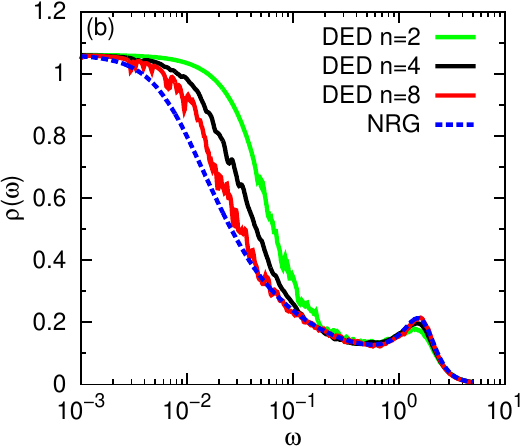} \\
    \includegraphics[width=0.48\linewidth]{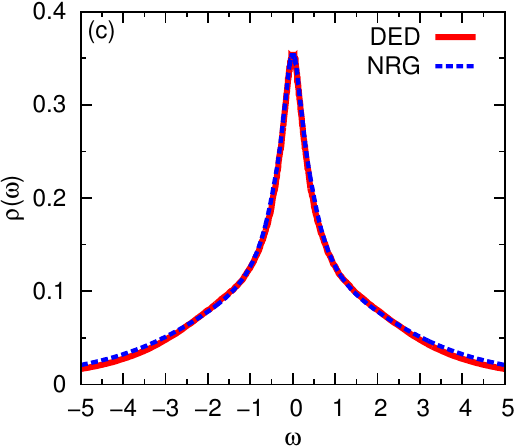} &
    \includegraphics[width=0.48\linewidth]{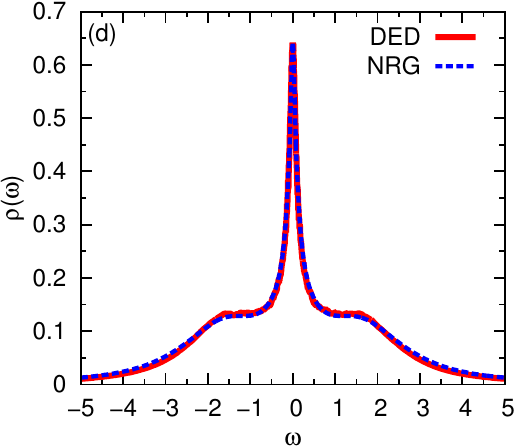} \\
    \includegraphics[width=0.48\linewidth]{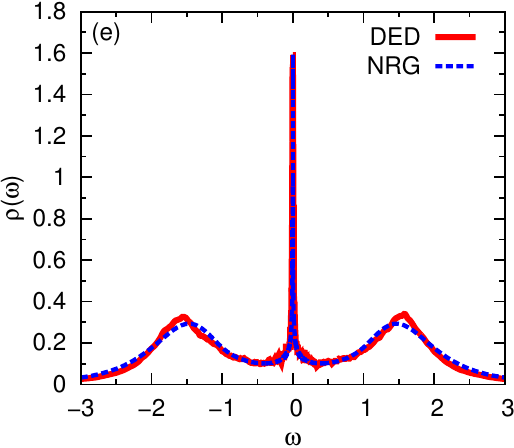} &
    \includegraphics[width=0.48\linewidth]{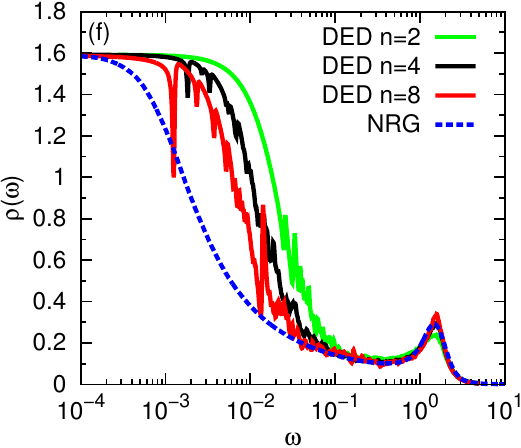} \\
    \includegraphics[width=0.48\linewidth]{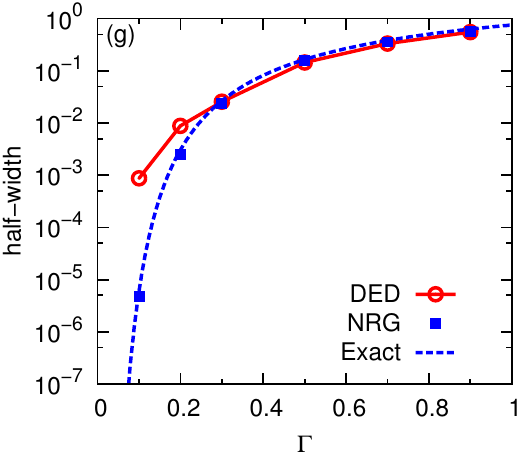} &
    \includegraphics[width=0.48\linewidth]{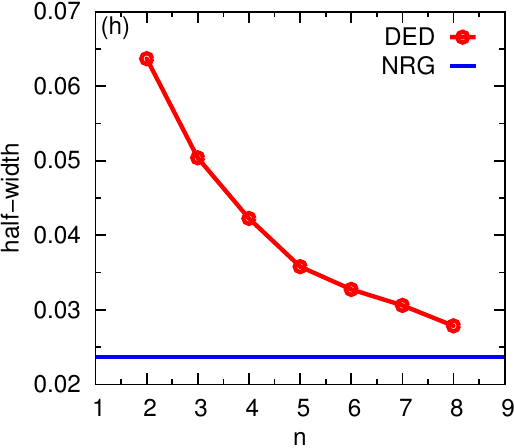}
  \end{tabular}
  \caption{\label{fig:SAIM}
   Comparison between DED and NRG spectra for the symmetric Anderson model. 
   ($U=3$, $\epsilon_d=-1.5$)
   (a) DED ($n=8$) and NRG spectra for $\Gamma=0.3$
   (b) NRG and DED spectra with different number of sites $n$ on a half-log scale for $\Gamma=0.3$.
   (c) DED ($n=6$) and NRG spectra for $\Gamma=0.9$.
   (d) DED ($n=6$) and NRG spectra for $\Gamma=0.5$. 
   (e) DED ($n=8$) and NRG spectra for $\Gamma=0.2$.
   (f) NRG and DED spectra for different number of sites $n$ on a half-log scale for $\Gamma=0.2$.
   (g) Half-width of Kondo peak estimated by fitting with Frota lineshapes\cite{Frota:PRB:1992} 
   versus $\Gamma$ calculated by DED ($n=8$ for $\Gamma\le0.3$ and $n=6$ for $\Gamma\ge0.5$) 
   compared to NRG and the exact expression\cite{Costi:JPCM:1994} on half-log scale.
   (h) Half-width of Kondo peak versus number of sites $n$ compared to NRG for $\Gamma=0.3$.
  }
 \end{figure}

First, we study the AIM at particle-hole symmetry, $\epsilon_d = -U/2$ and $\langle n_d \rangle=1$.
In this case the real part of the self-energy at the Fermi level is known prior to calculation,
$\Sigma_0=U/2$, and hence does not have to be determined self-consistently.
Fig.~\ref{fig:SAIM}a shows the impurity spectral function $\rho(\omega)=-\Im{G}(\omega)/\pi$ for 
$U=3$ and $\Gamma=0.3$ calculated by DED with $n=8$ sites, in comparison with the NRG spectrum. 
The DED and NRG spectra are in excellent overall agreement. 
The Anderson model is in the Kondo regime, where the spectral function is characterized by three 
resonances: The sharp Kondo resonance at the Fermi level and two Hubbard side peaks on either side 
of the Fermi level close to the excitation energies $\epsilon_d$ and $\epsilon_d+U$. 
In Fig.~\ref{fig:SAIM}b, we show DED spectra for different numbers of sites $n$
in comparison with NRG for the same set of parameters as in Fig.~\ref{fig:SAIM}a.
In order to better resolve the spectra at low energies, the energies are plotted on a logarithmic scale.
Even for very small models ($n=2$) there is good qualitative agreement with the NRG spectrum,
but the width of the Kondo peak is overestimated by a factor of almost 3 (see also Fig.~\ref{fig:SAIM}h),
and the height of the Hubbard side peaks is slightly underestimated. Note, however, that the height
of the Kondo peak $1/\pi \Gamma$ is always exact, independent of the number of sites $n$, since it is imposed 
by the particle constraint, as discussed in Sec.~\ref{sub_constraint}. As the number of sites $n$
increases, the quantitative agreement with NRG improves considerably, becoming excellent for $n=8$ sites. 
The quantitative improvement with increasing number of sites can also be seen in Fig.~\ref{fig:SAIM}h, where 
we show the half-width of the Kondo peak as a function of the model size $n$ in comparison to the 
NRG value. 

The number of randomly generated samples contributing to the ensemble average of the self-energy 
(\ref{eq:average}) generally determines the amount of noise in the spectra. For a fixed model size 
$n$, the noise can be reduced by increasing the number of samples $N$; it vanishes in the limit 
$N\rightarrow\infty$. On the other hand, the larger the number of sites $n$ of the finite size, 
the fewer samples are needed to achieve the same level of noise, since the number of poles
in the spectrum of individual samples increases. For example, in Fig.~\ref{fig:SAIM}b 
for $n=2,4,8$ sites about $5.8\cdot10^4$, $3.6\cdot10^4$ and $1.7\cdot10^4$ samples, respectively, were 
used to generate the spectra. In the limit $n\rightarrow\infty$ we would recover the continuous 
conduction band of the original Anderson model, and hence a single sample would already yield the 
exact and thus noiseless spectrum. In Tab.~\ref{tab:statistics} in App.~\ref{app_statistics} we 
report the number of samples used in calculating the spectra shown in Figs.~\ref{fig:SAIM} and \ref{fig:ASAIM}.

Next we investigate how the quality of the DED spectra changes when the correlation strength
controlled by $U/\Gamma$ is altered. In Fig.~\ref{fig:SAIM}c,d we show a comparison of spectra 
calculated by DED and NRG for higher values of the broadening $\Gamma$ than before. 
For weak correlation strength ($\Gamma=0.9$, Fig.~\ref{fig:SAIM}c), the system is no longer 
in the Kondo regime: the spectra are characterized by a single peak, though different from the 
Lorentzian of the non-interacting system due to interaction effects. Here the agreement with NRG 
is excellent already for $n=2$ (not shown). As the correlation strength increases, more sites are 
necessary to achieve good quantitative agreement. 
For $\Gamma=0.5$ (Fig.~\ref{fig:SAIM}d), we approach the Kondo regime, and the three peak structure
starts to emerge. Now excellent quantitative agreement with NRG can be achieved for $n=6$ sites.
We have already discussed the case $\Gamma=0.3$ (Figs.~\ref{fig:SAIM}a,b,h), already in the Kondo regime,
where excellent agreement with NRG is reached for $n=8$ sites.
Figures~\ref{fig:SAIM}e,f show DED spectra in comparison with NRG for $\Gamma=0.2$, on a normal energy
scale (Fig.~\ref{fig:SAIM}e), and on a logarithmic energy scale (Fig.~\ref{fig:SAIM}f) for better resolution of 
the low-energy features. The overall qualitative agreement with the NRG spectrum is again quite good,
as can be seen from Fig.~\ref{fig:SAIM}e. However, the quantitative agreement, especially of the low
energy features, i.e. the Kondo peak, is not very good anymore: the width of the Kondo peak is still
strongly overestimated by almost a factor of 2 even for $n=8$ sites. The high energy features on the 
other hand are captured quite well, although the height of the Hubbard side peaks is slightly overestimated.

This behavior of decreasing quality of the DED at a fixed number of sites with increasing correlation 
strength $U/\Gamma$ is summarized in Fig.~\ref{fig:SAIM}g which shows the half-width of the Kondo peak 
as a function of $\Gamma$, comparing DED for $n=8$ sites and NRG. For not too strong correlations, i.e. 
$\Gamma\ge0.3$ ($U/\Gamma\le10$), DED for $n=8$ sites yields an excellent approximation to the width of 
the Kondo peak, but begins to deviate from NRG as the correlations become stronger (decreasing $\Gamma$). 
For very strong correlations (i.e. $U/\Gamma\gg10$), the width of the Kondo peak becomes strongly 
overestimated, by orders of magnitude (see also Fig.~\ref{fig:SAIM}e and Fig.~\ref{fig:SAIM}f).  
This behavior can be understood by considering the Kondo screening cloud, whose spatial extension 
grows exponentially with increasing correlation strength:\cite{Affleck:2002} $\xi_K\propto1/T_K\propto{e^{\Gamma/U}}$.
Thus the number of bath sites necessary to correctly describe the Kondo screening cloud 
grows exponentially with the correlation strength, leading generally to an overestimation of
the Kondo temperature for too small bath sizes.
Hence for very strong correlation strengths the DED method cannot provide a quantitatively
satisfactory description of the spectra with a computationally feasible number of bath sites.
Yet for correlation strengths up to and including $U/\Gamma\approx10$ DED yields an excellent description 
of the spectra for small to moderate numbers of bath sites.

\subsection{Asymmetric Anderson model}
\label{sub_asymm_aim} 

\begin{figure}
  \begin{tabular}{cc}
    \includegraphics[width=0.48\linewidth]{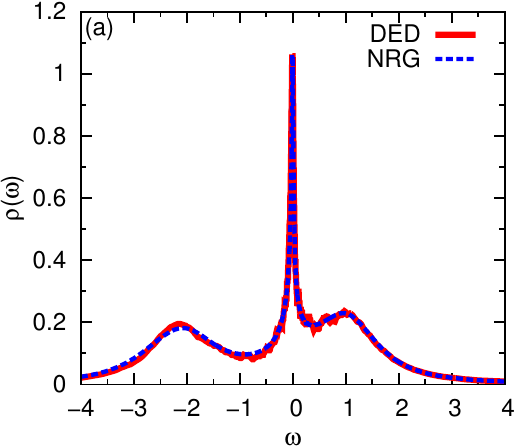} &
    \includegraphics[width=0.48\linewidth]{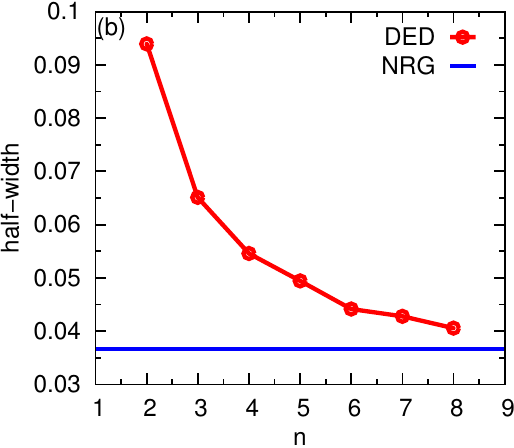} \\
    \multicolumn{2}{c}{\includegraphics[width=0.96\linewidth]{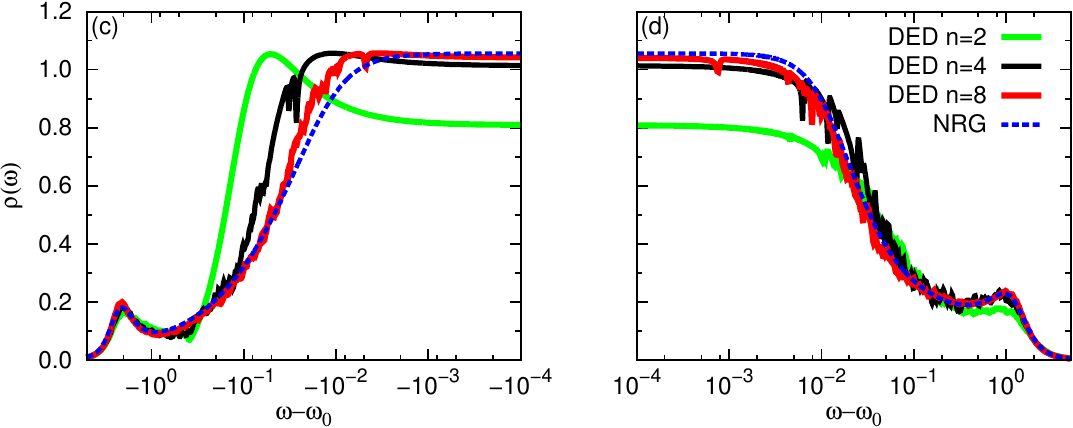}} \\
    \includegraphics[width=0.48\linewidth]{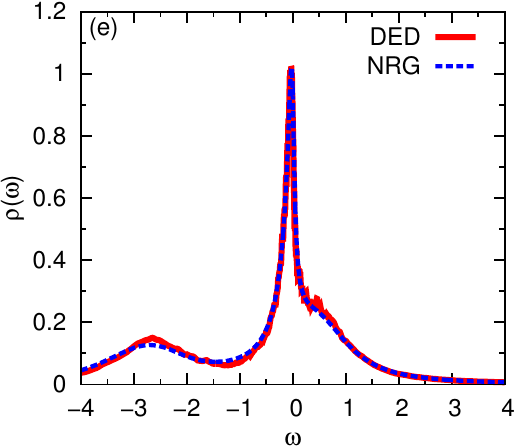} &
    \includegraphics[width=0.48\linewidth]{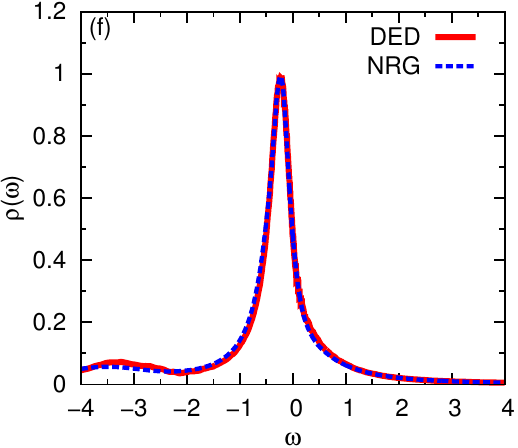}
  \end{tabular}
  \caption{\label{fig:ASAIM}
    Comparison between DED and NRG spectra for the asymmetric Anderson model ($U=3$, $\epsilon_d<-1.5$, $\Gamma=0.3$). 
    (a) DED (for $n=8$ sites) and NRG spectra for $\epsilon_d=-2$.
    (b) Half-width of Kondo peak versus number of sites $n$ compared to NRG ($\epsilon_d=-2$).
    (c) NRG and DED spectra with different number of sites $n$ for $\epsilon_d=-2$ on a half-log scale for negative energies.
    (d) NRG and DED spectra with different number of sites $n$ for $\epsilon_d=-2$ on a half-log scale for positive energies.
    (e) DED (for $n=8$ sites) and NRG spectra for $\epsilon_d=-2.5$.
    (e) DED (for $n=8$ sites) and NRG spectra for $\epsilon_d=-3$.
  }
\end{figure}

We now consider the AIM away from particle-hole symmetry, $\epsilon_d<-U/2 $ and $\langle n_d \rangle>1$. 
As explained before in Sec.~\ref{sub_constraint}, the parameter $\Sigma_0$ has 
to be determined self-consistently since we identified it with the real part of the 
self-energy at the Fermi level, $\Sigma_0\equiv\Re\bar\Sigma(\ef)$ which is unknown prior
to the DED calculation. We thus start with some reasonable initial guess, for example
the Hartree shift $\Sigma_0\equiv{U\langle{n_d}\rangle/2}$, calculated within Hartree-Fock, 
$\Sigma_0\equiv{U}/2$, or simply $\Sigma_0\equiv-\epsilon_d$. Using this $\Sigma_0$ in the DED procedure we calculate
the self-energy $\bar\Sigma^{(1)}(\omega)$ and thus obtain a new guess for 
$\Sigma_0\rightarrow\Re\bar\Sigma^{(1)}(\ef)$, and repeat until self-consistency is reached. 
This procedure usually converges within a few cycles (3-4) to an accuracy of under 1\%.
We find that the effect of the self-consistency on the overall spectrum is relatively small. 
The main effect is to improve the position of the Kondo peak and to recover the exact 
height of the Kondo peak. Hence if the fine details of the spectrum 
are less important, it suffices to compute $\bar\Sigma(\omega)$ for some reasonable 
guess, for example $\Sigma_0=U/2$. More details on the self-consistent determination
of $\Sigma_0$ can be found in App.~\ref{app_sigma0}.

In Fig.~\ref{fig:ASAIM}a we show the impurity spectral density for $U=3$, $\Gamma=0.3$ 
and $\epsilon_d=-2$, calculated by DED for $n=8$ sites in comparison with the NRG spectrum. 
The DED spectrum is in very good overall agreement with the NRG spectrum. 
For as much as we are in the Kondo regime the three peak structure is retained (see Sec.~\ref{sub_symm_aim}).
As in the symmetric case, in order to better resolve the spectra at low energies, we use a logarithmic scale
for the energy axis. Since here we are dealing with asymmetric spectra, we represent the spectral density on 
the logarithmic scale for negative and positive energies in Fig.~\ref{fig:ASAIM}c and Fig.~\ref{fig:ASAIM}d, 
respectively. 
As in the ph symmetric case, we observe quantitative improvement of the DED spectra with increasing $n$.
Especially the position and width of the Kondo peak improve considerably: While for small $n$ the peak 
is considerably offset from the Fermi level, the pinning of the Kondo peak to the Fermi level as seen 
in NRG is almost completely recovered for $n=8$. As can be seen from Fig.~\ref{fig:ASAIM}b, similar to 
the symmetric case (see Sec.~\ref{sub_symm_aim}), the width of the Kondo peak is strongly overestimated 
for $n=2$ by almost a factor of 3, but decreases rapidly with increasing $n$, until for $n=8$ the width
is only slightly overestimated by a few percent.

Next we investigate the quality of the DED spectra when moving away from the Kondo regime, by further 
decreasing $\epsilon_d$ such that $\epsilon_d+U$ approaches the Fermi level. In Figs.~\ref{fig:ASAIM}e,f
we compare spectra calculated by DED (for $n=8$ sites) and by NRG in the intermediate valence regime 
$\epsilon_d+U-\epsilon_F\approx\Gamma$. In this regime the charge of the impurity level fluctuates 
strongly between single and double occupation, leading to a significant deviation of $\langle{n_d}\rangle$
from unity. The spectral density is characterized by two resonances, one at $\approx\epsilon_d+U$ of width 
$\approx\Gamma$, and a much less pronounced resonance at $\epsilon_d$. Upon further decreasing $\epsilon_d$
the resonance at $\epsilon_d$ becomes more strongly suppressed [compare Fig.~\ref{fig:ASAIM}f with 
Fig.~\ref{fig:ASAIM}e], as we get closer to the non-magnetic regime ($\epsilon_d+U-\epsilon_F\ll-\Gamma$) 
where the impurity level is almost doubly occupied, and the resonance finally vanishes (not shown).
As can be seen from Figs.~\ref{fig:ASAIM}e,f, the DED spectra are in excellent agreement with the NRG 
ones even for strong asymmetry, capturing all the described features very well.

\begin{table}
  \begin{tabular*}{\linewidth}{@{\extracolsep{\fill}}lcccc}
    \hline
    $\epsilon_d$ & NRG & DED (ENS) & DED (FSR) & $\Sigma_0$ \\
    \hline
    -1.5  & 1.0000 & 0.9992$\pm$0.0062 & 1.008$\pm$0.068 & 1.496$\pm$0.032 \\
    -1.65 & 1.0202 & 1.0234$\pm$0.0063 & 1.039$\pm$0.064 & 1.630$\pm$0.030 \\
    -1.8  & 1.0420 & 1.0495$\pm$0.0066 & 1.065$\pm$0.057 & 1.769$\pm$0.027 \\
    -2.0  & 1.0765 & 1.0862$\pm$0.0089 & 1.132$\pm$0.056 & 1.937$\pm$0.027 \\
    -2.5  & 1.2322 & 1.2366$\pm$0.0091 & 1.309$\pm$0.035 & 2.341$\pm$0.017 \\
    -3.0  & 1.5270 & 1.5364$\pm$0.0128 & 1.533$\pm$0.023 & 2.667$\pm$0.012 \\
    \hline
  \end{tabular*}
  \caption{
    \label{tab:occ}
    The $d$-level occupancy $\langle{n_d}\rangle$ calculated by NRG compared to DED 
    obtained (i) via the ensemble average (ENS) and (ii) via Friedel's sum rule (FSR) from 
    $\Re\bar\Sigma(\ef)$ as well as the self-consistently determined $\Sigma_0=\Re\bar\Sigma(\ef)$ 
    for different values of $\epsilon_d$ and their statistical errors.\cite{Note2}
  }
\end {table}

Finally, we also calculate the occupancy of the impurity level $\langle{n_d}\rangle$ 
for different values of $\epsilon_d$ and compare with NRG. We investigate two different
ways of calculating $\langle n_d\rangle$ within DED. On the one hand we can calculate
the occupancy from the ensemble average (ENS) over accepted finite Anderson model samples:
\begin{equation}
  \label{eq:ENS}
  \langle n_d \rangle \approx \bar{n}_d = \frac{1}{N} \sum_\nu \bra{0^\nu} n_d \ket{0^\nu}
\end{equation}
On the other hand we can make use of Friedel's sum rule (FSR), and calculate $\langle{n_d}\rangle$ 
from the self-energy at the Fermi level:
\begin{equation}
  \label{eq:FSR}
  \langle n_d  \rangle = 1 - \frac{2}{\pi} \tan^{-1}\left( \frac{\epsilon_d+\Re\bar\Sigma(\ef)-\ef}{\Gamma} \right) 
\end{equation}
where we have already taken into account spin-degeneracy. Also note that 
$n_{{\rm imp},\sigma}= \langle{n_{d,\sigma}}\rangle$ in the flat wide band limit.\cite{Hewson:book:1997}
Table~\ref{tab:occ} shows the results for NRG and DED using $n=8$ sites. The overall
agreement between DED and NRG is very good.
The values of $\langle{n_d}\rangle$ calculated by both approaches in DED
agree with the NRG results within the statistical error\footnote{ 
  The statistical errors of the ensemble averaged occupancy and of $\Sigma_0$
  were estimated from the standard deviation from the mean over all accepted samples.
  In the case of the occupancy calculated from $\Sigma_0$ by FSR, the statistical
  error was calculated by error propagation from the standard deviation of $\Sigma_0$,
  i.e. $\delta{n}_d = |\partial{n}_d/\partial{\Sigma_0}|_{\bar\Sigma_0} \delta{\Sigma}_0$
}
for all values of $\epsilon_d$. However, the statistical error is generally
smaller for the ENS approach. Only for very strong asymmetry ($\epsilon_d=-3$) does 
the error of the FSR approach become slightly smaller than the one of the ENS approach,
and also the mean values are closer to the NRG results for ENS than for FSR.

From Tab.~\ref{tab:occ} we can see that the error in the occupancy $\langle{n_d}\rangle$ 
calculated via FSR as well as the error in $\Sigma_0$ decrease with increasing asymmetry. 
This can be understood by considering the acceptance ratio which becomes better the 
stronger the asymmetry (see App.~\ref{app_statistics}) so that more samples are accepted (for a fixed 
total number of samples) contributing to the ensemble average for the self-energy, and thus improving 
the statistics. For small asymmetries the argument to $\tan^{-1}$ in FSR (\ref{eq:FSR}) is close to zero
($\epsilon_d+\Re\bar\Sigma(\ef)-\ef\approx0$)), and therefore $\tan^{-1}$ has an approximately
linear behavior so that $\delta{n_d}\approx\frac{2}{\pi\Gamma}\delta\Sigma_0$, explaining
the factor of roughly two between the error in $\langle{n_d}\rangle$ and the error 
in $\Sigma_0$ since $2/\pi\Gamma\approx2$ for $\Gamma=0.3$. 
%%For larger asymmetries ($\epsilon_d\le-2.5$), the sub-linear behavior of $\tan^{-1}$ 
%%leads to much more stronger decrease for $\delta{n_d}$ than for $\delta\Sigma_0$.
On the other hand, the error for $\langle{n_d}\rangle$ calculated via ENS \emph{increases}
slightly with increasing asymmetry despite more samples being accepted, since the 
occupancies of individual finite Anderson model samples fluctuate more strongly with
increasing asymmetry.

\section{Conclusions}
\label{sec_conclusions}

In conclusion, we find that DED generally yields an excellent description of the Anderson impurity model, 
inside as well as outside the Kondo regime. The spectra obtained by DED are in good qualitative 
agreement with NRG spectra already for a small number of bath sites. Depending on the correlation strength
$U/\Gamma$ excellent quantitative agreement can be achieved for a moderate number of 5-7 bath sites.
Only for very strong correlation, $U/\Gamma\gg10$, does the number of bath sites necessary to achieve
a good quantitative description become too big to be computationally feasible due to the exponential growth
of the Kondo screening cloud.

We further find that the particle number constraint plays an essential role in the DED method for the
description of Kondo physics. Basically, the constraint ensures that individual finite Anderson model 
samples contributing to the self-energy average comply with Nozieres' Fermi liquid picture of the 
strong coupling regime, thereby imposing Fermi liquid behavior on the sample averaged self-energy.
This leads to the recovery of the Kondo peak in the spectrum, which is absent in the DED procedure without 
the constraint. 

The enforcement of Fermi liquid behavior by the constraint means that its role needs to be 
reconsidered when DED is applied to situations where Fermi liquid behavior is not obeyed, for example, 
at finite temperatures above $T_K$, in gapped systems, or in the case of multi-orbital Anderson models 
where non-Fermi liquid behavior may occur.\cite{Nozieres:JP:1980,Schlottmann:AP:1993,DeLeo:PRB:2005}
More precisely, it seems that the constraint needs to be relaxed in some way in order to describe the 
loss of Fermi liquid behavior in these cases. As can be seen from Fig.~\ref{fig:constraint}a, without 
the constraint DED produces a spectrum similar to that of the Anderson model in the Coulomb blockade regime. 
In other words, DED with the constraint describes the strong coupling fixed point of the Anderson model, while
DED without the constraint seems to describe the weak coupling fixed point. 
This may also explain why a DED+DMFT scheme without application of the constraint is capable of describing the 
gapped Mott insulating phase of the Hubbard model.\cite{Granath:PRB:2014} 
Thus in order to describe the transition from the Fermi liquid to the Coulomb blockade or Mott regime
a general principle for relaxing the constraint needs to be found.

An advantage of DED over NRG is that it can be parallelized very efficiently 
as the randomly generated finite Anderson model samples can be diagonalized independently from each other, 
and hence can be easily distributed over an arbitrary large number of nodes. This recommends DED for the 
solution of multi-orbital Anderson models which are not accessible for NRG for more than three impurity levels. 
Adopting the Lanczos diagonalization scheme in the DED procedure should allow one to treat multi-orbital Anderson 
models with a sufficient number of bath sites per impurity level to achieve accurate results. Another 
advantage of DED is that the energy resolution is the same on all energy scales and thus can be exploited in resolving 
higher energy spectral features that would be difficult to resolve with NRG.\cite{Granath:PRB:2014}

\begin{appendix}

\section{The constraint and the 1:1 correspondence with the non-interacting system}
\label{app_constraint}

\begin{figure}[b]
  \includegraphics[width=0.98\linewidth]{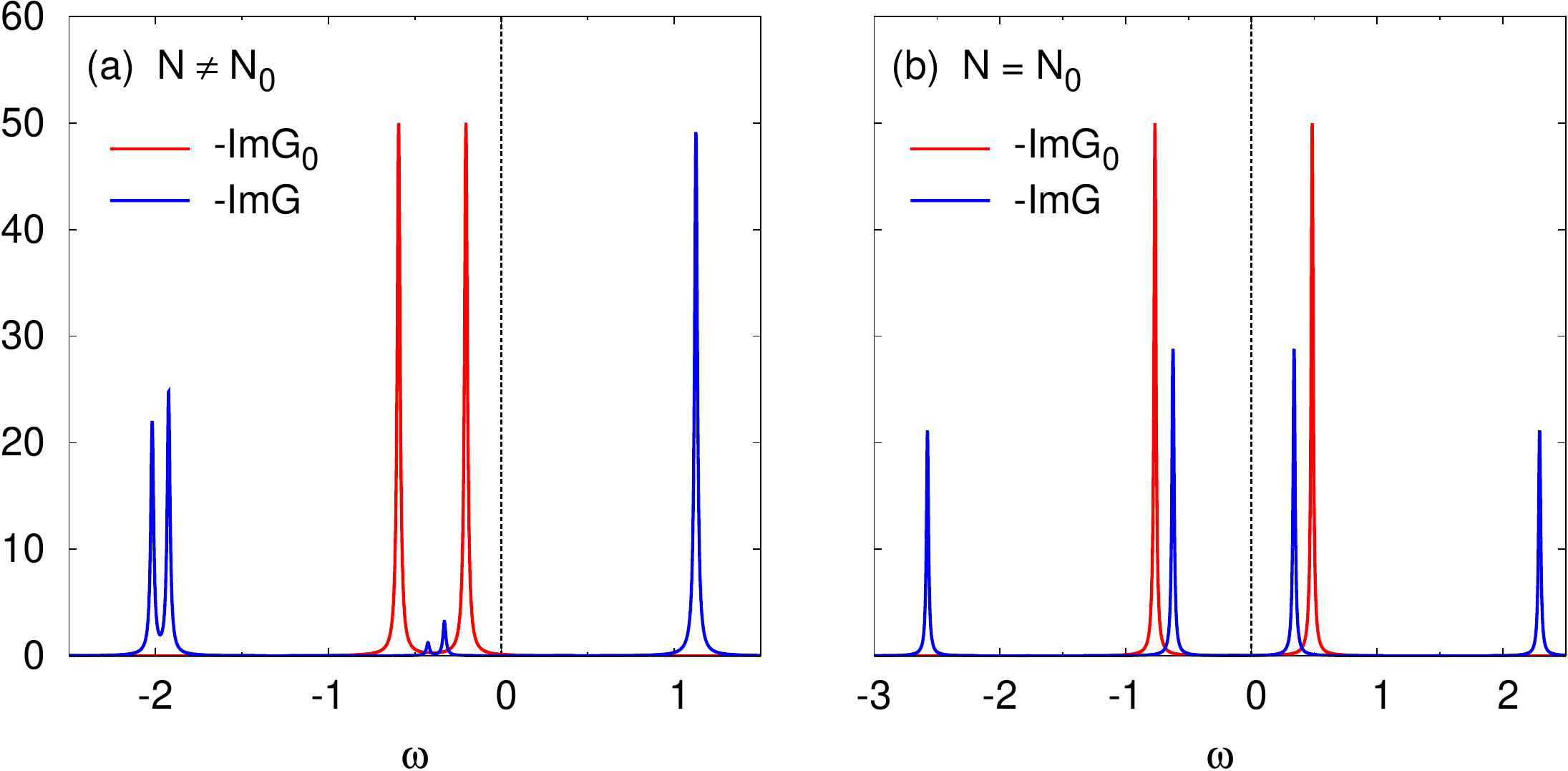}
  \caption{
    \label{fig:sample}
    Comparison of interacting (blue) and non-interacting (red) finite Anderson model spectra
    ($n=2$ sites) for the case that (a) the particle constraint is not fulfilled, and (b) when it is fulfilled.
    Only in the latter case a 1:1 correspondence between the interacting and non-interacting
    system can be established.
  }
\end{figure}

As discussed in Sec.~\ref{sub_constraint}, the particle number constraint (\ref{eq:constraint})
ensures that every finite Anderson model sample contributing to the self-energy average (\ref{eq:average}) 
obeys Fermi liquid behavior, i.e. requires that a 1:1 correspondence can be established between
the interacting model and the corresponding non-interacting effective model. In the following we discuss 
in more detail how this 1:1 correspondence is established via the constraint. 

First, note that since each Anderson model sample $\nu$ is finite, the interacting and non-interacting 
particle numbers $n_{\rm imp}^\nu$ and $n_{{\rm imp},0}^\nu$, respectively, are discrete (integer) numbers, 
and thus also the corresponding phase shifts $\eta_\sigma^\nu(\ef)$ and $\eta_{0,\sigma}^\nu(\ef)$ are discrete numbers.
Depending on the signs of the numerators in the arguments to $\tan^{-1}$ in eqs. (\ref{eq:eta}) and (\ref{eq:eta0}), 
the phase shifts can assume either the value $0$ (negative) or $\pi$ (positive), since for a finite
system generally $\Im\Delta^\nu(\omega)\rightarrow0$ as $\delta\rightarrow0^+$ (unless $\omega$ is at 
a pole), and hence the argument to $\tan^{-1}$ diverges, i.e. goes to $\pm\infty$ depending on the sign of the 
numerator. A phase shift of $\pi/2$ is theoretically also possible (implying $n_{\rm imp}=1$), but in practice does not happen, 
as it means that either a bath level $k$ is exactly at the Fermi level ($\epsilon^\nu_k=\ef$), 
so that $\Im\Delta^\nu(\ef)\rightarrow\infty$ as $\delta\rightarrow0^+$, or the numerator is 
exactly zero, meaning that the sampled poles lie exactly symmetric w.r.t. the Fermi level.
Hence during the DED procedure the phase shift of individual samples will fluctuate between 
the two values $0$ and $\pi$. In the ph symmetric situation ($\epsilon_d=-U/2$) the number of samples
with phase shift 0 will be equal to the number of samples with phase shift $\pi$ for a large
enough number of samples. Hence on average we obtain the phase shift of $\pi/2$. Away from ph symmetry,
the number of samples with one phase shift grows at the expense of samples with the other phase shift,
leading to an average phase shift different from $\pi/2$.

The sign of the numerators $\ef-\epsilon_d^\nu-\Re\Sigma^\nu_\sigma(\ef)$ in (\ref{eq:eta})
and $\ef-\epsilon_d^\nu-\Sigma_0$ in (\ref{eq:eta0}) are largely determined by the positions
of the most important excitations with respect to the Fermi level. If the most important 
excitation is hole-like, then the numerator is negative and hence the phase shift is 0. If
on the contrary the excitation is electron-like, the numerator is positive and hence leads to
a phase shift of $\pi$. Therefore the constraint is only fulfilled (i.e. matching phase shifts of
interacting and corresponding non-interacting system) if the most important excitation in the 
interacting and non-interacting system are of the same type, i.e. either both hole-like or both 
electron-like. This is illustrated in Fig.~\ref{fig:sample} which compares the spectra 
of an interacting and non-interacting finite Anderson model in the case that the constraint is not 
fulfilled (a) and when it is fulfilled (b). One can clearly see that the main excitations 
are not of the same type when the constraint is not fulfilled, while they are of the same type
if the constraint is fulfilled. Clearly, in the latter case a 1:1 correspondence can be established
between the excitations of the interacting and corresponding non-interacting system.

\section{DED Statistics}
\label{app_statistics}

\begin{table}[h]
  \begin{tabular*}{\linewidth}{@{\extracolsep{\fill}}lccrrc}
    \hline
    $\epsilon_d$ & $\Gamma$ & $n$ & $N_{\rm tot}$ & $N_{\rm acc}$ & $r_{\rm acc}$ \\
    \hline
    -1.5         & 0.2      & 8   & 23875       & 7449        & 31\%  \\
    -1.5         & 0.3      & 2   & 200000      & 57561       & 29\%  \\
    -1.5         & 0.3      & 4   & 100000      & 35918       & 36\%  \\
    -1.5         & 0.3      & 8   & 38594       & 16504       & 43\%  \\
    -1.5         & 0.5      & 6   & 100000      & 55066       & 55\%  \\
    -1.5         & 0.9      & 6   & 100000      & 70297       & 70\%  \\
    -1.65        & 0.3      & 8   & 8500        & 3743        & 44\%  \\
    -1.8         & 0.3      & 8   & 8495        & 3847        & 45\%  \\
    -2.0         & 0.3      & 8   & 8495        & 4037        & 47\%  \\
    -2.5         & 0.3      & 8   & 7958        & 4809        & 60\%  \\
    -3.0         & 0.3      & 8   & 8000        & 6452        & 81\%  \\
    \hline
  \end{tabular*}
  \caption{
    \label{tab:statistics}
    Summary of statistical information of the DED calculations reported in the 
    text. Total number of samples $N_{\rm tot}$, number of accepted samples $N_{\rm acc}$,
    and the acceptance ratio $r_{\rm acc}$ for different values of $\epsilon_d$, $\Gamma$ 
    and number of sites $n$. For all calculations $U=3$ was used. 
  }
\end{table}

In Table~\ref{tab:statistics} we summarize statistical information on the DED 
calculations reported in the text. One can see that the acceptance ratio $r_{\rm acc}$ 
increases with increasing $\Gamma$ (i.e. decreasing correlation strength $U/\Gamma$),
and increasing asymmetry. In both cases interaction effects become weaker, 
so that the non-interacting limit is approached, where the DED becomes
exact already for the one-site model (the non-interacting DOS can
be reproduced by simply sampling the non-interacting DOS $\rho_0(\omega)$
of course) where the constraint is always fulfilled.

\section{Self-consistent determination of $\Sigma_0$}
\label{app_sigma0}

\begin{figure}[b]
  \includegraphics[width=\linewidth]{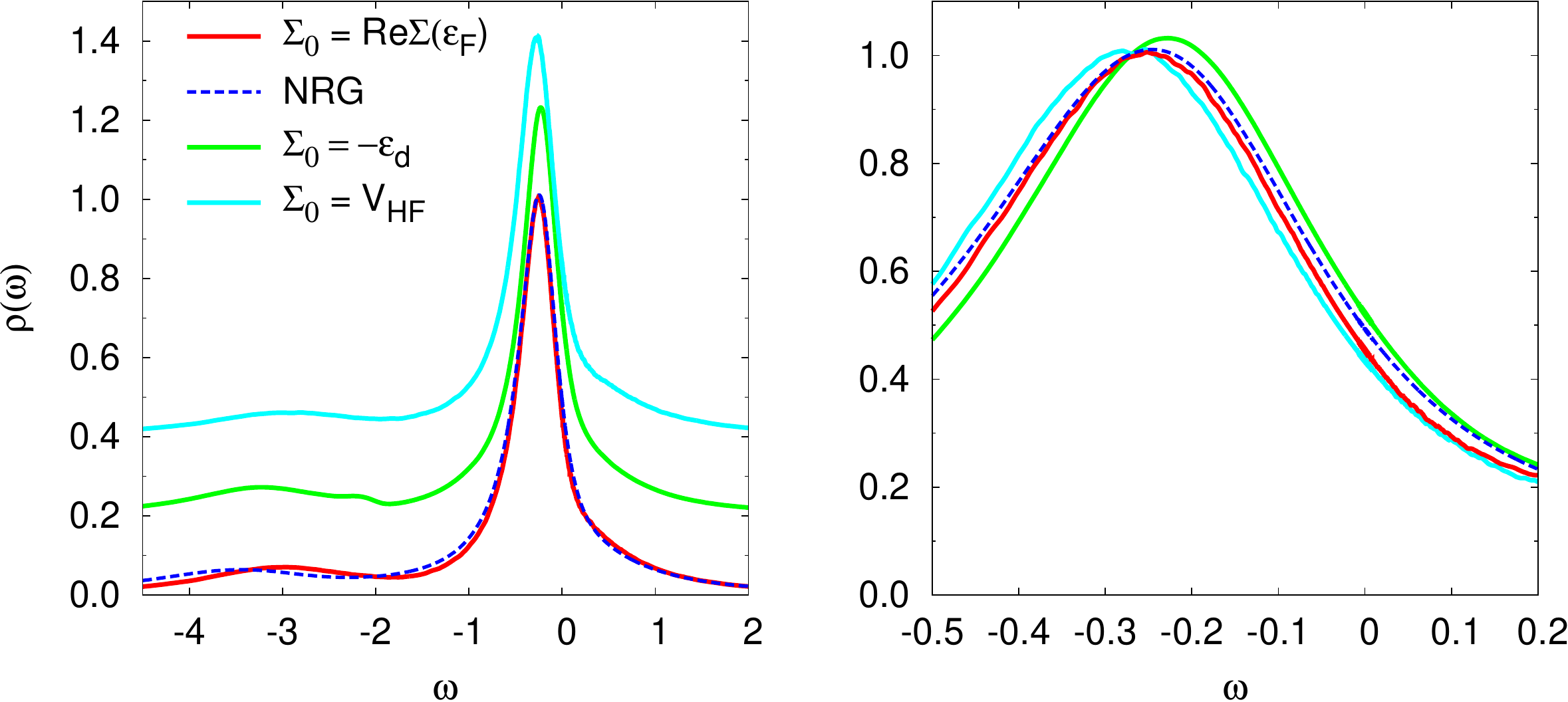}
  \caption{\label{fig:sigma0}
    Spectral function close to the Fermi level for different values of $\Sigma_0$ calculated by DED
    ($n=4$ sites) compared to NRG for $U=3$, $\epsilon_d=-3$ and $\Gamma=0.3$ on a large energy
    scale (left) and at low energies (right). Note that on the left panel the curves for $\Sigma_0=-\epsilon_d$
    (green) and for $\Sigma_0=V_{\rm HF}$ (cyan) have been offset by 0.2 and 0.4, respectively, 
    in order to increase the visibility. 
  }
\end{figure}

As explained in Sec.~\ref{sub_constraint} the effective one-body potential $\Sigma_0$ 
entering the non-interacting GF (\ref{eq:GF0}) should be identified with the real part
of the self-energy at the Fermi level, $\Sigma_0\equiv\Re\Sigma(\ef)$. However, with the 
exception of the ph symmetric situation where $\Re\Sigma(\ef)=U/2$, the self-energy at the
Fermi level is unknown prior to calculation. 
Hence we propose to determine $\Sigma_0$ self-consistently, by starting with some 
initial guess, e.g. $\Sigma_0\rightarrow{U/2}$. Using this initial guess the DED procedure
yields $\Re\bar\Sigma(\ef)$, generally different from $\Sigma_0$, which is taken
as the new guess, $\Sigma_0\rightarrow\Re\bar\Sigma(\ef)$. This procedure is repeated 
until self-consistency is reached, i.e. $\Sigma_0$ does not change anymore within a 
specified accuracy. We find that the self-consistency converges quite rapidly to an accuracy
of under 1\% within 3-4 cycles. In Fig.~\ref{fig:sigma0} we show the effect of the 
self-consistency for $\Sigma_0$ on the spectra close to the Fermi level. 
The agreement between NRG and DED using the converged value $\Sigma_0=\Re\bar\Sigma(\ef)\approx2.65$ 
(red line) is quite good.
But the effect of self-consistency is actually relatively weak: DED with the initial guess 
$\Sigma_0=-\epsilon_d$ or using the Hartree shift for $\Sigma_0$ yield spectra that 
are also quite close to the NRG spectrum, with the peak position just slightly shifted,
even when using the Hartree-Fock potential $\Sigma_0\equiv{V_{\rm HF}}\approx2.49$ (cyan line).

\end{appendix}

\bibliography{nanodmft}

\end{document}